\documentclass[reprent,aps,prb,twocolumn,
superscriptaddress,nofootinbib]{revtex4-2}
\usepackage{amsmath,amssymb}
\usepackage{graphicx}
\usepackage{bm}
\usepackage{multirow}
\usepackage{hyperref}
\usepackage{braket}
\usepackage{mathrsfs}

\usepackage{color}

\allowdisplaybreaks[2]

\begin{document}

\title{
Classification of multipoles induced by external fields and currents\\ under electronic nematic ordering with quadrupole moments
}
\author{Satoru Hayami}
\email{hayami@phys.sci.hokudai.ac.jp}
\affiliation{Graduate School of Science, Hokkaido University, Sapporo 060-0810, Japan}

\begin{abstract}
We theoretically investigate the effect of external fields and currents on electronic nematic orderings based on the concept of augmented multipoles consisting of electric, magnetic, magnetic toroidal, and electric toroidal multipoles. 
We show the relation between rank-2 electric quadrupoles and the other multipoles, the former of which corresponds to the microscopic order parameter for the nematic phases. 
The electric (magnetic) field induces the rank-1 and rank-3 electric (magnetic) multipoles and rank-2 electric toroidal (magnetic toroidal) quadrupoles, while the electric current induces the rank-1 and rank-3 magnetic toroidal multipoles and rank-2 magnetic quadrupoles. 
We classify the active multipoles under magnetic point groups, which will be a reference to explore cross-correlation and transport phenomena in nematic phases. 
\end{abstract}

\maketitle

\section{Introduction}

The orbital degree of freedom in electrons has attracted much interest in condensed matter physics, since it becomes a source of unconventional electronic orderings and their related physical phenomena~\cite{tokura2000orbital, streltsov2017orbital, khomskii2021orbital}. 
Among them, electronic nematic orderings, which appear through the spontaneous breaking of rotational symmetry in solids, have been extensively studied in both theory and experiments. 
In contrast to magnetic orderings, the breaking of time-reversal ($\mathcal{T}$) symmetry is not necessary, and hence, qualitatively distinct low-energy excitations and physical phenomena are expected. 
The electronic nematic states have been discussed in various contexts, such as the Pomeranchuk instability~\cite{pomeranchuk1959sov, Yamase_JPSJ.69.332, Halboth_PhysRevLett.85.5162}, spin nematic state~\cite{Fath_PhysRevB.51.3620, Lauchli_PhysRevB.74.144426, Manmana_PhysRevB.83.184433, tsunetsugu2006spin, Nic_PhysRevLett.96.027213, Shindou_PhysRevB.80.064410, tsunetsugu2007spin, Smerald_PhysRevB.88.184430}, and charge/orbital nematic state~\cite{kivelson1998electronic, Fradkin_PhysRevB.59.8065, Emery_PhysRevLett.85.2160, Kee_PhysRevB.71.184402, chuang2010nematic, Goto_JPSJ.80.073702, yoshizawa2012structural}, which have been observed in $d$-electron materials like Ba$_2$MgReO$_6$~\cite{Hirai_PhysRevResearch.2.022063, Mansouri_PhysRevMaterials.5.104410, Lovesey_PhysRevB.103.235160} and $f$-electron materials, such as CeB$_6$~\cite{Takigawa_doi:10.1143/JPSJ.52.728,nakao2001antiferro,tanaka2004direct,Portnichenko_PhysRevX.10.021010}, PrPb$_3$~\cite{morin1982magnetic, onimaru2004angle, Onimaru_PhysRevLett.94.197201}, Pr$T_2$$X_{20}$ ($T=$ Ir, Rh, $X=$ Zn; $T=$ V, $X=$ Al)~\cite{Onimaru_PhysRevLett.106.177001, Ishii_doi:10.1143/JPSJ.80.093601, sakai2011kondo, onimaru2016exotic, Ishitobi_PhysRevB.104.L241110}, and CeCoSi~\cite{tanida2018substitution, tanida2019successive, yatsushiro2020odd, Manago_doi:10.7566/JPSJ.90.023702, Hidaka_doi:10.7566/JPSJ.91.094701, Matsumura_doi:10.7566/JPSJ.91.064704, Manago_PhysRevB.108.085118}.
Recently, further exotic states coexisting spin and nematic orders have been suggested, such as the CP$^2$ skyrmion~\cite{Garaud_PhysRevB.87.014507, Akagi_PhysRevD.103.065008, Amari_PhysRevB.106.L100406, zhang2023cp2} and other multiple-$Q$ states~\cite{tsunetsugu2021quadrupole, Hattori_PhysRevB.107.205126, Ishitobi_PhysRevB.107.104413,hayami2023cluster, hayami2023multiple}.

\begin{figure}[tb!]
\begin{center}
\includegraphics[width=1.0\hsize]{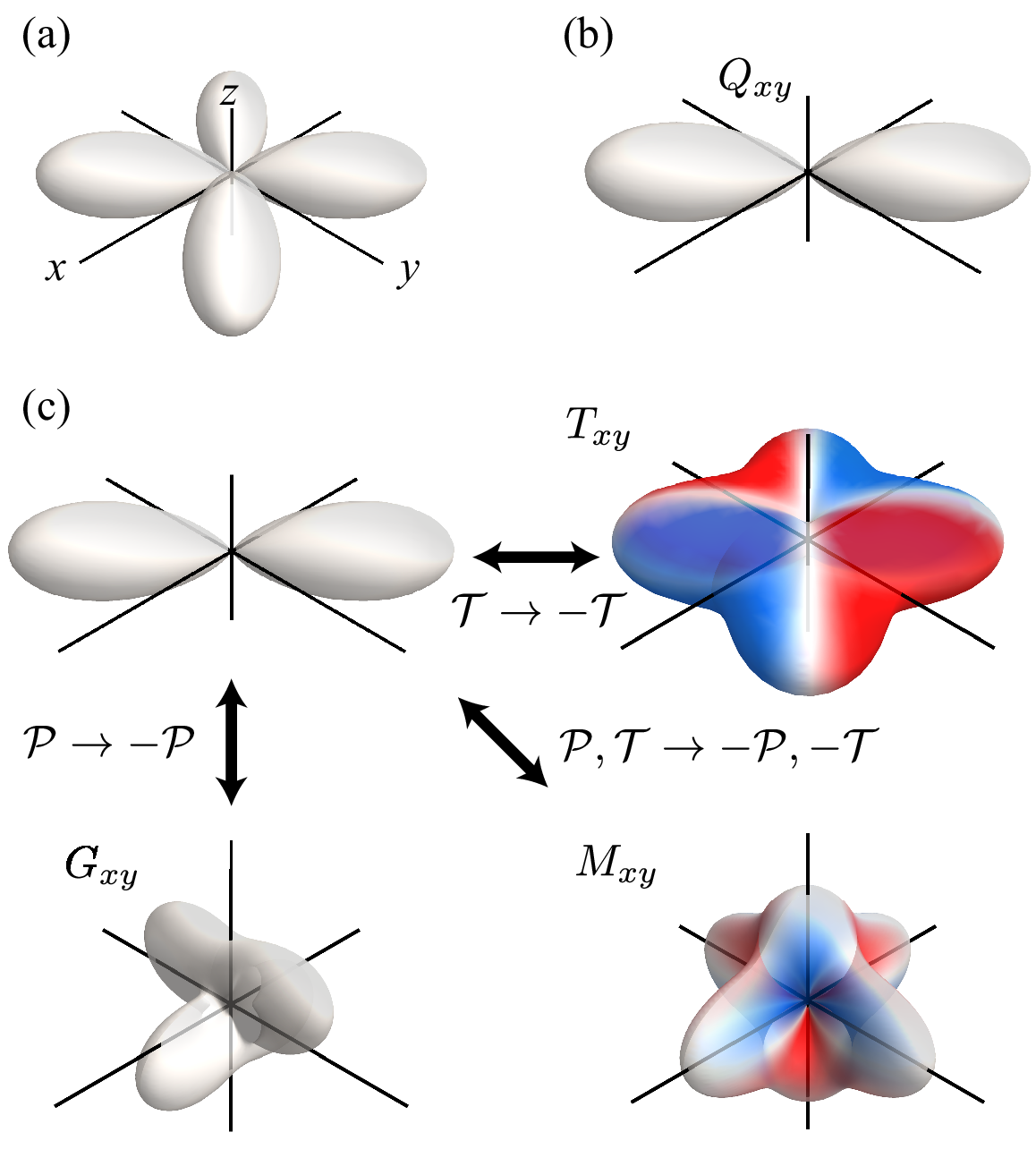} 
\caption{
\label{fig: ponti} 
(a) Example of the orbital state under the tetragonal symmetry, where its shape represents the spatial charge distribution. 
(b) The charge distribution of the electric quadrupole $Q_{xy}$, which appears as a result of the breaking of fourfold rotational symmetry. 
(c) The conversion of the spatial inversion ($\mathcal{P}$) and time-reversal ($\mathcal{T}$) parities for $Q_{xy}$; the opposite $\mathcal{P}$ ($\mathcal{T}$) parity leads to the electric toroidal quadrupole $G_{xy}$ (magnetic toroidal quadrupole $T_{xy}$), and the opposite $\mathcal{P}$ and $\mathcal{T}$ lead to the magnetic quadrupole $M_{xy}$. 
The red and blue colors represent the $z$ component of the positive and negative orbital-angular momentum density, respectively. 
}
\end{center}
\end{figure}

The breaking of the rotational symmetry is related to the emergence of multipole moments, since the multipoles of rank 1 or higher describe the spatial anisotropy. 
The microscopic order parameter in the nematic state can be described by the multipole moment; the electric multipole characterized by the polar tensor with time-reversal even corresponds to the order parameter. 
For example, when the fourfold rotational symmetry of the tetragonal system in Fig.~\ref{fig: ponti}(a) is broken, the order parameter is expressed as the $xy$ component of the electric quadrupole, $Q_{xy} \propto xy$, where the spatial charge distribution is shown in Fig.~\ref{fig: ponti}(b).  
Meanwhile, such an electric quadrupole breaks neither the spatial inversion ($\mathcal{P}$) symmetry nor $\mathcal{T}$ symmetry. 

In the present study, we investigate the effect of the breakings of $\mathcal{P}$ and $\mathcal{T}$ symmetries under the electronic nematic orderings accompanying the electric quadrupole. 
Especially, we focus on further symmetry breakings by external fields and currents. 
Based on symmetry and microscopic multipole representation analyses~\cite{hayami2018microscopic, Hayami_PhysRevB.98.165110, kusunose2022generalization}, we show the coupling between rank-2 electric quadrupole and external stimuli. 
As a result, we find that external stimuli induce various multipole moments through $\mathcal{P}$ and/or $\mathcal{T}$ symmetry breakings; the electric (magnetic) field induces the electric (magnetic) dipoles and octupoles, and electric toroidal (magnetic toroidal) quadrupole, while the electric current induces the magnetic toroidal dipoles and octupoles, and magnetic quadrupoles. 
We discuss the relevant cross-correlation and transport properties in each case. 
Furthermore, we classify the symmetry lowering under these stimuli in terms of active multipoles. 
Our systematic investigation will provide the possibility of not only acquiring functionalities related to the violation of $\mathcal{P}$ and $\mathcal{T}$ parities but also generating and controlling multipole moments under nematic orderings. 

The rest of this paper is organized as follows. 
In Sec.~\ref{sec: Active multipoles under external fields and currents}, we show active multipoles induced by external fields and currents under electronic nematic orderings. 
Then, we classify such field-induced multipoles under magnetic point groups in Sec.~\ref{sec: Classification of field-induced multipoles under point group}. 
Section~\ref{sec: Summary} is devoted to a summary of this paper.

\section{Active multipoles under external fields and currents}
\label{sec: Active multipoles under external fields and currents}

In this section, we show what types of multipoles emerge by applying the external fields and currents to the nematic phases. 
For that purpose, let us introduce four types of multipoles for later discussion. 
Four types of multipoles correspond to the electric multipole $Q_{lm}$ with the $\mathcal{P}$ and $\mathcal{T}$ parities of $(\mathcal{P}, \mathcal{T})=[(-1)^l, +1]$, magnetic multipole $M_{lm}$ with $(\mathcal{P}, \mathcal{T})=[(-1)^{l+1}, -1]$, magnetic toroidal multipole $T_{lm}$ with $(\mathcal{P}, \mathcal{T})=[(-1)^l, -1]$, and electric toroidal multipole $G_{lm}$ with $(\mathcal{P}, \mathcal{T})=[(-1)^{l+1}, +1]$, where $l$ and $m$ represent the rank of multipoles and its component. 
In terms of the $\mathcal{P}$ and $\mathcal{T}$ parities, the electric multipole $Q_{lm}$, which are described by the spherical harmonics, is related to $G_{lm}$ by reversing $\mathcal{P}$, $T_{lm}$ by reversing $\mathcal{T}$, and $M_{lm}$ by reversing both $\mathcal{P}$ and $\mathcal{T}$~\cite{hayami2016emergent, hayami2018microscopic, Hayami_PhysRevB.98.165110}. 
For example, the counterparts of the electric quadrupole $Q_{xy}$ in the other three multipoles are given by the electric toroidal quadrupole $G_{xy}$, magnetic toroidal quadrupole $T_{xy}$, and magnetic quadrupole $M_{xy}$, respectively, whose spatial distributions in terms of the charge and orbital-angular momentum densities are schematically shown in Fig.~\ref{fig: ponti}(c)~\cite{hayami2018microscopic}; it is noted that the $\mathcal{P}$ ($\mathcal{T}$) symmetry is lost in $G_{xy}$ ($T_{xy}$), while both $\mathcal{P}$ and $\mathcal{T}$ symmetries are lost in $M_{xy}$. 
These four types of multipole constitute a complete set in the Hilbert space in electron systems~\cite{kusunose2020complete, Kusunose_PhysRevB.107.195118}. 

Among these multipoles, the electric quadrupole $Q_{2m}=(Q_u, Q_v, Q_{yz}, Q_{zx}, Q_{xy}) \propto (3z^2-r^2, x^2-y^2, yz, zx, xy)$ under invariant $\mathcal{P}$ and $\mathcal{T}$ corresponds to the order parameter under electronic nematic orderings. 
In other words, the rotational symmetry breaking and its related band deformation under electronic nematic orderings, which have been studied in iron-based superconductors~\cite{Fang_PhysRevB.77.224509, Fernandes_PhysRevLett.105.157003, Kruger_PhysRevB.79.054504, Lv_PhysRevB.80.224506, Lee_PhysRevLett.103.267001, Onari_PhysRevLett.109.137001} and Sr$_3$Ru$_2$O$_7$~\cite{Raghu_PhysRevB.79.214402, Lee_PhysRevB.80.104438, Tsuchiizu_PhysRevLett.111.057003}, are attributed to the appearance of $Q_{2m}$.

In the following, we show the coupling between the electric quadrupole and other multipoles via external fields and currents. 
First, we show the specific expressions of their coupling in Sec.~\ref{sec: Coupling to vector fields}. 
Then, we apply the results for the electric field in Sec.~\ref{sec: Electric field}, the magnetic field in Sec.~\ref{sec: Magnetic field}, and the electric current in Sec.~\ref{sec: Electric current}.

\subsection{Coupling to vector fields}
\label{sec: Coupling to vector fields}

We consider the coupling between the electric quadrupole $Q_{2m}$ and the external vector fields/currents $\bm{F}=(F_x, F_y, F_z)$ based on the group theory. 
Since $Q_{2m}$ and $\bm{F}$ correspond to the rank-2 and rank-1 quantities, their product leads to the rank-1 dipole, rank-2 quadrupole, and rank-3 octupole quantities, which are denoted by $(\mathcal{D}_{x}, \mathcal{D}_{y}, \mathcal{D}_{z})$, $(\mathcal{Q}_u, \mathcal{Q}_v, \mathcal{Q}_{yz}, \mathcal{Q}_{zx}, \mathcal{Q}_{xy})$, and $(\mathcal{O}_{xyz}, \mathcal{O}^\alpha_{x}, \mathcal{O}^\alpha_{y}, \mathcal{O}^\alpha_{z}, \mathcal{O}^\beta_{x}, \mathcal{O}^\beta_{y}, \mathcal{O}^\beta_{z})$, respectively. 

The rank-1 dipole-type coupling is given by 
\begin{align}
\label{eq:Dx}
\mathcal{D}_x&= \left(-\frac{1}{\sqrt{3}}Q_u + Q_v\right) F_x+Q_{xy} F_y + Q_{zx} F_z, \\
\label{eq:Dy}
\mathcal{D}_y &=Q_{xy}F_x -\left(\frac{1}{\sqrt{3}}Q_u + Q_v \right)F_y+Q_{yz} F_z , \\
\label{eq:Dz}
\mathcal{D}_z &=Q_{zx} F_x + Q_{yz}F_y+\frac{2}{\sqrt{3}}Q_u F_z,
\end{align}
where $(\mathcal{D}_x, \mathcal{D}_y, \mathcal{D}_y)$ represent the nonuniform field distinct from the uniform field $(F_x, F_y, F_z)$, which is referred to as an anisotropic dipole~\cite{Hayami_PhysRevB.103.L180407}. 
The rank-2 quadrupole-type coupling is given by 
\begin{align}
\label{eq:Qu}
\mathcal{Q}_{u}&=\sqrt{3} Q_{yz} F_x - \sqrt{3} Q_{zx} F_y, \\
\label{eq:Qv}
\mathcal{Q}_v &= Q_{yz} F_x + Q_{zx} F_y - 2 Q_{xy} F_z, \\
\label{eq:Qyz}
\mathcal{Q}_{yz} &= -(\sqrt{3} Q_u + Q_v) F_x - Q_{xy} F_y + Q_{zx} F_z, \\
\label{eq:Qzx}
\mathcal{Q}_{zx} &= Q_{xy} F_x -(-\sqrt{3} Q_u + Q_v) F_y  - Q_{yz} F_z, \\
\label{eq:Qxy}
\mathcal{Q}_{xy} &= -Q_{zx} F_x + Q_{yz} F_y + 2Q_{v} F_z, 
\end{align}
and the rank-3 octupole-type coupling is given by 
\begin{align}
\label{eq:Oxyz}
\mathcal{O}_{xyz}&= Q_{yz} F_x +  Q_{zx} F_y + Q_{xy} F_z, \\
\label{eq:Oxa}
\mathcal{O}^{\alpha}_x &=\sqrt{\frac{3}{5}} \left[ \frac{\sqrt{3}}{2} (-Q_u + \sqrt{3} Q_v) F_x -Q_{xy} F_y -Q_{zx} F_z \right], \\
\label{eq:Oya}
\mathcal{O}^{\alpha}_{y} &= \sqrt{\frac{3}{5}} \left[ -Q_{xy} F_x -\frac{\sqrt{3}}{2} (Q_u + \sqrt{3} Q_v) F_y -Q_{yz} F_z \right], \\
\label{eq:Oza}
\mathcal{O}^{\alpha}_{z} &= \sqrt{\frac{3}{5}} \left[ -Q_{zx} F_x - Q_{yz} F_y+ \sqrt{3} Q_u F_z \right], \\
\label{eq:Oxb}
\mathcal{O}^{\beta}_{x} &= -\frac{1}{2}(\sqrt{3} Q_u + Q_v)F_x + Q_{xy} F_y -Q_{zx} F_z, \\
\label{eq:Oyb}
\mathcal{O}^{\beta}_{y} &= -Q_{xy} F_x +\frac{1}{2} (\sqrt{3} Q_u - Q_v) F_y+ Q_{yz} F_z, \\
\label{eq:Ozb}
\mathcal{O}^{\beta}_{z} &= Q_{zx} F_x - Q_{yz} F_y + Q_{v} F_z, 
\end{align}
where the overall numerical coefficient is appropriately taken in each rank for simplicity. 
$\mathcal{D}$, $\mathcal{Q}$, and $\mathcal{O}$ denote any of electric multipole $Q$, magnetic multipole $M$, magnetic toroidal multipole $T$, and electric toroidal multipole $G$, which depends on the $\mathcal{P}$ and $\mathcal{T}$ parities in $\bm{F}$.

\begin{table}[tb!]
\begin{center}
\caption{
The relationship between the external vector fields $\bm{F}=(F_x, F_y, F_z)$ and the rank-1--3 multipoles under the electric quadrupole (EQ) $Q_{2m}=(Q_u, Q_v, Q_{yz}, Q_{zx}, Q_{xy})$. 
}
\label{tab: mp}
\begingroup
\renewcommand{\arraystretch}{1.1}
\scalebox{1.0}{
 \begin{tabular}{lccc}
 \hline \hline
EQ & $F_x$ & $F_y$ & $F_z$ \\ \hline
$Q_u$ & ($\mathcal{D}_x$, $\mathcal{Q}_{yz}$, $\mathcal{O}^{\alpha,\beta}_x$) & ($\mathcal{D}_y$, $\mathcal{Q}_{zx}$, $\mathcal{O}^{\alpha, \beta}_{y}$) & ($\mathcal{D}_z$, $\mathcal{O}^{\alpha}_{z}$)  \\
$Q_v$ & ($\mathcal{D}_x$, $\mathcal{Q}_{yz}$, $\mathcal{O}^{\alpha, \beta}_x$) & ($\mathcal{D}_y$, $\mathcal{Q}_{zx}$, $\mathcal{O}^{\alpha,\beta}_{y}$) & ($\mathcal{Q}_{xy}$, $\mathcal{O}^{\beta}_{z}$)  \\
$Q_{yz}$ & ($\mathcal{Q}_u$, $\mathcal{Q}_v$, $\mathcal{O}_{xyz}$) & ($\mathcal{D}_z$, $\mathcal{Q}_{xy}$, $\mathcal{O}^{\alpha,\beta}_{z}$) & ($\mathcal{D}_y$, $\mathcal{Q}_{zx}$, $\mathcal{O}^{\alpha,\beta}_{y}$)  \\
$Q_{zx}$ & ($\mathcal{D}_z$, $\mathcal{Q}_{xy}$, $\mathcal{O}^{\alpha,\beta}_{z}$) & ($\mathcal{Q}_u$, $\mathcal{Q}_v$, $\mathcal{O}_{xyz}$) & ($\mathcal{D}_x$, $\mathcal{Q}_{yz}$, $\mathcal{O}^{\alpha,\beta}_x$)  \\
$Q_{xy}$ & ($\mathcal{D}_y$, $\mathcal{Q}_{zx}$, $\mathcal{O}^{\alpha,\beta}_{y}$) & ($\mathcal{D}_x$, $\mathcal{Q}_{yz}$, $\mathcal{O}^{\alpha,\beta}_x$) & ($\mathcal{Q}_v$, $\mathcal{O}_{xyz}$)   \\
\hline\hline  
\end{tabular}
}
\endgroup
\end{center}
\end{table}

From these couplings, one finds what types of multipole degrees of freedom are induced when the external fields and currents are applied under electronic nematic orderings. 
The different types of multipoles can be induced depending on the types of electric quadrupoles and field direction. 
For example, when the expectation value of $Q_{xy}$ is nonzero, the rank-1 dipole $\mathcal{D}_y$, the rank-2 quadrupole $\mathcal{Q}_{zx}$, and the rank-3 octupoles $(\mathcal{O}^{\alpha}_{y}, \mathcal{O}^{\beta}_y)$ are induced by the $x$-directional field $F_x$, while the rank-2 quadrupole $\mathcal{Q}_v$ and the rank-3 octupole $\mathcal{O}_{xyz}$ are induced by the $z$-directional field $F_z$. 
We summarize the relationship between $\bm{F}$ and induced multipoles under $Q_{2m}$ in Table~\ref{tab: mp}. 
We discuss the specific multipoles induced by external fields and currents in the following subsections.

\subsection{Electric field}
\label{sec: Electric field}

When $\bm{F}$ is the electric field $\bm{E}$ with the spatial inversion and time-reversal parities of $(\mathcal{P}, \mathcal{T})=(-1,+1)$, $\mathcal{D}$, $\mathcal{Q}$, and $\mathcal{O}$ correspond to the electric dipole $Q_{1m}$, electric toroidal quadrupole $G_{2m}$, and electric octupole $Q_{3m}$, respectively. 
We discuss several characteristic features by applying $\bm{E}$. 

For the dipole component $Q_{1m}$, the transverse electric polarization appears for $(Q_{yz}, Q_{zx}, Q_{xy})$ in addition to the conventional longitudinal one.
For example, the $y$($x$)-directional electric polarization related to $Q_y$ ($Q_x$) appears for $Q_{xy}$ under $E_x$ ($E_y$). 
It is noted that this transverse response is symmetric by interchanging $E_x$ and $E_y$, which is in contrast to the antisymmetric rotational response under the ferroaxial ordering; $Q_y$ ($Q_x$) is induced by $E_x$ ($E_y$) for the $Q_{xy}$ ordering, while $Q_y$ ($-Q_x$) is induced by $E_x$ ($E_y$) for the ferroaxial ordering~\cite{Hayami_doi:10.7566/JPSJ.91.113702, cheong2022linking, kirikoshi2023rotational}. 
Since the electric dipole $Q_{1m}$ leads to the antisymmetric spin splitting in the electronic band structure, the Edelstein effect, where the magnetization is induced by the electric current, can be expected. 

Depending on $Q_{2m}$ and $\bm{E}$, the dipole component vanishes, but the quadrupole component is finite, as shown in Table~\ref{tab: mp}. 
One of the examples is the situation where $E_z$ is applied to the $Q_{xy}$ state; the electric toroidal quadrupole $G_v$ is induced. 
This result indicates that the electric field becomes a conjugate field of the electric toroidal quadrupole under the electronic nematic orderings. 
Thus, physical phenomena brought about by the electric toroidal quadrupole, such as the nonlinear Hall effect based on the Berry curvature dipole mechanism~\cite{Sodemann_PhysRevLett.115.216806} and Edelstein effect discussed in materials hosting the electric toroidal quadrupole~\cite{Matteo_PhysRevB.96.115156, Hayami_PhysRevLett.122.147602, Ishitobi_doi:10.7566/JPSJ.88.063708, hayami2023chiral}, are expected.

\subsection{Magnetic field}
\label{sec: Magnetic field}

In the case of the magnetic field $\bm{F}=\bm{H}$ with $(\mathcal{P}, \mathcal{T})=(+1,-1)$, the relevant multipoles are rank-1 magnetic dipole $M_{1m}$, rank-2 magnetic toroidal quadrupole $T_{2m}$, and rank-3 magnetic octupole $M_{3m}$. 
Similarly to the electric field, the transverse magnetization related to the magnetic dipole $(M_x, M_y, M_z)$ is expected for the orderings with $(Q_{yz}, Q_{zx}, Q_{xy})$. 
In addition, the magnetic toroidal quadrupole $T_{2m}$ is induced depending on $Q_{2m}$ and $\bm{H}$. 
Since the magnetic toroidal quadrupole becomes the origin of the directional-dependent spin current generation arising from the symmetric-type momentum-dependent spin splitting in the band structure~\cite{hayami2022spinconductivity, Hayami_PhysRevB.108.L140409}, similar phenomena are also expected when the magnetic field is applied to the electronic nematic orderings. 
Moreover, the rank-3 magnetic octupole is induced through the coupling between $Q_{2m}$ and $\bm{H}$, which has been proposed and observed in CeB$_6$ hosting the electric quadrupole ordered state~\cite{shiina1997magnetic, sakai1997new, shiina1998interplay, shiba1999nature, sakai1999antiferro, Matsumura_PhysRevLett.103.017203, Matsumura_PhysRevB.85.174417}.

\subsection{Electric current}
\label{sec: Electric current}

Finally, we consider the case of the electric current, i.e., $\bm{F}=\bm{J}$. 
Since the electric current corresponds to the polar vector with time-reversal odd to satisfy $(\mathcal{P}, \mathcal{T})=(-1,-1)$, the corresponding multipoles to $\mathcal{D}$, $\mathcal{Q}$, and $\mathcal{O}$ are the magnetic toroidal dipole $T_{1m}$, magnetic quadrupole $M_{2m}$, and magnetic toroidal octupole $T_{3m}$, respectively, where the magnetic toroidal dipole becomes the origin of the linear antisymmetric magneto-electric effect~\cite{Spaldin_0953-8984-20-43-434203, kopaev2009toroidal, Yanase_JPSJ.83.014703, Hayami_doi:10.7566/JPSJ.84.064717} and nonreciprocal transport~\cite{Watanabe_PhysRevResearch.2.043081, Yatsushiro_PhysRevB.105.155157}, the magnetic quadrupole leads to the linear symmetric magneto-electric effect, current-induced distortion, and nonlinear Hall effect~\cite{Watanabe_PhysRevB.96.064432, thole2018magnetoelectric, Yanagi_PhysRevB.97.020404, Hayami_PhysRevB.97.024414, Hayami_PhysRevB.104.045117, Hayami_PhysRevB.106.024405, Kirikoshi_PhysRevB.107.155109}, and magnetic toroidal octupole gives rise to nonreciprocal transport owing to the asymmetric band structure~\cite{Hayami_doi:10.7566/JPSJ.85.053705}.

\begin{table}[tb!]
\begin{center}
\caption{
The correspondence between external fields/currents and induced multipoles. 
}
\label{tab: mp_field}
\begingroup
\renewcommand{\arraystretch}{1.1}
\scalebox{1.0}{
 \begin{tabular}{ccccc}
 \hline \hline
$\bm{F}=(F_x, F_y, F_z)$ & $\mathcal{D}$ (rank 1) & $\mathcal{Q}$ (rank 2) & $\mathcal{O}$ (rank 3) \\ \hline
Electric field $\bm{E}$ &  $Q_{1m}$ &  $G_{2m}$  &  $Q_{3m}$  \\
Magnetic field $\bm{H}$ &  $M_{1m}$ &  $T_{2m}$  &  $M_{3m}$  \\
Electric current $\bm{J}$ &  $T_{1m}$ &  $M_{2m}$  &  $T_{3m}$  \\
\hline\hline  
\end{tabular}
}
\endgroup
\end{center}
\end{table}

In this way, electronic nematic orderings acquire various functionalities according to the emergence of multipoles when the external fields and currents are applied. 
We summarize the relationship between external fields/currents and induced multipoles in Table~\ref{tab: mp_field}. 
It is noted that the above results can be applied to the coupling to other vector fields and currents. 
For example, the spin current in the form of $\bm{J} \times \bm{\sigma}$ ($\bm{\sigma}$ is the spin polarization vector) is categorized into the case of $\bm{F}=\bm{E}$. 
When $\bm{F}$ corresponds to the time-reversal-even axial vectors, such as the static rotational distortion, the corresponding multipoles are given by the electric toroidal dipole for $\mathcal{D}$, electric quadrupole for $\mathcal{Q}$, and electric toroidal octupole for $\mathcal{O}$.

\section{Classification of field-induced multipoles under point group}
\label{sec: Classification of field-induced multipoles under point group}

\begin{table}[tb!]
\begin{center}
\caption{
Multipoles induced by the electric field $\bm{E}=(E_x, E_y, E_z)$, the magnetic field $\bm{H}=(H_x, H_y, H_z)$, and the electric current $\bm{J}=(J_x, J_y, J_z)$ under the tetragonal magnetic point group (MPG) $4/mmm 1'$. 
We omit the multipoles for the electric monopole and the electric toroidal octupole for simplicity. 
}
\label{tab: D4h}
\begingroup
\renewcommand{\arraystretch}{1.1}
\scalebox{1.0}{
 \begin{tabular}{lcc}
 \hline \hline
$\bm{F}$ & MPG & multipoles \\ \hline
 ---      & $4/mmm 1'$ & $Q_u$ \\
 \hline
$E_x$ & $2mm 1'$ & $Q_x$, $Q_v$, $Q^\alpha_x$, $Q^\beta_x$, $G_{yz}$  \\
$E_y$ & $m2m 1'$ & $Q_y$, $Q_v$, $Q^\alpha_y$, $Q^\beta_y$, $G_{zx}$\\
$E_z$ & $4mm 1'$ & $Q_z$, $Q^\alpha_z$\\
\hline
$H_x$ & $mm'm'$ & $Q_v$, $M_x$, $M^{\alpha}_x$, $M^\beta_{x}$, $T_{yz}$ \\
$H_y$ & $m'mm'$ & $Q_v$, $M_y$, $M^{\alpha}_y$, $M^\beta_{y}$, $T_{zx}$ \\
$H_z$ & $4/mm'm'$ & $M_z$, $M^\alpha_z$  \\
\hline
$J_x$ & $m'mm$ & $Q_v$, $T_x$, $T^{\alpha}_x$, $T^\beta_{x}$, $M_{yz}$ \\
$J_y$ & $mm'm$ & $Q_v$, $T_y$, $T^{\alpha}_y$, $T^\beta_{y}$, $M_{zx}$  \\
$J_z$ & $4/m'mm$ & $T_z$, $T^\alpha_z$   \\
\hline\hline  
\end{tabular}
}
\endgroup
\end{center}
\end{table}

\begin{table}[tb!]
\begin{center}
\caption{
Multipoles induced by $\bm{E}=(E_x, E_y, E_z)$, $\bm{H}=(H_x, H_y, H_z)$, and $\bm{J}=(J_x, J_y, J_z)$ under the hexagonal magnetic point group (MPG) $6/mmm 1'$. 
We omit the multipoles for the electric monopole and the electric toroidal octupole for simplicity. 
}
\label{tab: D6h}
\begingroup
\renewcommand{\arraystretch}{1.1}
\scalebox{1.0}{
 \begin{tabular}{lcc}
 \hline \hline
$\bm{F}$ & MPG & multipoles \\ \hline
 ---      & $6/mmm 1'$ & $Q_u$ \\
 \hline
$E_x$ & $2mm 1'$ & $Q_x$, $Q_v$, $Q^\alpha_x$, $Q^\beta_x$, $G_{yz}$  \\
$E_y$ & $m2m 1'$ & $Q_y$, $Q_v$, $Q^\alpha_y$, $Q^\beta_y$, $G_{zx}$\\
$E_z$ & $6mm 1'$ & $Q_z$, $Q^\alpha_z$\\
\hline
$H_x$ & $mm'm'$ & $Q_v$, $M_x$, $M^{\alpha}_x$, $M^\beta_{x}$, $T_{yz}$ \\
$H_y$ & $m'mm'$ & $Q_v$, $M_y$, $M^{\alpha}_y$, $M^\beta_{y}$, $T_{zx}$ \\
$H_z$ & $6/mm'm'$ & $M_z$, $M^\alpha_z$  \\
\hline
$J_x$ & $m'mm$ & $Q_v$, $T_x$, $T^{\alpha}_x$, $T^\beta_{x}$, $M_{yz}$ \\
$J_y$ & $mm'm$ & $Q_v$, $T_y$, $T^{\alpha}_y$, $T^\beta_{y}$, $M_{zx}$  \\
$J_z$ & $6/m'mm$ & $T_z$, $T^\alpha_z$   \\
\hline\hline  
\end{tabular}
}
\endgroup
\end{center}
\end{table}

\begin{table}[tb!]
\begin{center}
\caption{
Multipoles induced by $\bm{E}=(E_x, E_y, E_z)$, $\bm{H}=(H_x, H_y, H_z)$, and $\bm{J}=(J_x, J_y, J_z)$ under the trigonal magnetic point group (MPG) $\bar{3}m 1'$. 
We omit the multipoles for the electric monopole, the electric toroidal dipole, and the electric toroidal octupole for simplicity. 
}
\label{tab: D3d}
\begingroup
\renewcommand{\arraystretch}{1.1}
\scalebox{1.0}{
 \begin{tabular}{lcc}
 \hline \hline
$\bm{F}$ & MPG & multipoles \\ \hline
 ---      & $\bar{3}m 1'$ & $Q_u$ \\
 \hline
$E_x$ & $m 1'$ & $Q_x$, $Q_z$, $Q_v$, $Q_{zx}$, $Q^\alpha_x$, $Q^\alpha_z$, $Q^\beta_x$, $Q^\beta_z$, $G_{yz}$, $G_{xy}$ \\
$E_y$ & $2 1'$ & $Q_y$, $Q_v$, $Q_{zx}$, $Q_{xyz}$, $Q^\alpha_y$, $Q^\beta_y$, $G_0$, $G_u$, $G_v$, $G_{zx}$ \\
$E_z$ & $3m 1'$ & $Q_z$, $Q^\alpha_z$, $Q_{3a}$\\
\hline
$H_x$ & $2'/m'$ & $Q_v$, $Q_{zx}$, $M_{x}$, $M_{z}$, $M^\alpha_x$, $M^\alpha_z$, $M^\beta_x$, $M^\beta_z$, $T_{yz}$, $T_{xy}$ \\
$H_y$ & $2/m$ & $Q_v$, $Q_{zx}$, $M_y$, $M_{xyz}$, $M^\alpha_y$, $M^\beta_y$, $T_0$, $T_u$, $T_v$, $T_{zx}$ \\
$H_z$ & $\bar{3}m'$ & $M_z$, $M^\alpha_z$, $M_{3a}$  \\
\hline
$J_x$ & $2'/m$ & $Q_v$, $Q_{zx}$, $T_{x}$, $T_{z}$, $T^\alpha_x$, $T^\alpha_z$, $T^\beta_x$, $T^\beta_z$, $M_{yz}$, $M_{xy}$ \\
$J_y$ & $2/m'$ & $Q_v$, $Q_{zx}$, $T_y$, $T_{xyz}$, $T^\alpha_y$, $T^\beta_y$, $M_0$, $M_u$, $M_v$, $M_{zx}$  \\
$J_z$ & $\bar{3}'m$ & $T_z$, $T^\alpha_z$, $T_{3a}$   \\
\hline\hline  
\end{tabular}
}
\endgroup
\end{center}
\end{table}

\begin{table}[tb!]
\begin{center}
\caption{
Multipoles induced by $\bm{E}=(E_x, E_y, E_z)$, $\bm{H}=(H_x, H_y, H_z)$, and $\bm{J}=(J_x, J_y, J_z)$ under the orthorhombic magnetic point group (MPG) $mmm 1'$. 
We omit the multipoles for the electric monopole and the electric toroidal octupole for simplicity. 
}
\label{tab: D2h}
\begingroup
\renewcommand{\arraystretch}{1.1}
\scalebox{1.0}{
 \begin{tabular}{lcc}
 \hline \hline
$\bm{F}$ & MPG & multipoles \\ \hline
 ---      & $mmm 1'$ & $Q_u$, $Q_v$ \\
 \hline
$E_x$ & $2mm 1'$ & $Q_x$, $Q^\alpha_x$, $Q^\beta_x$, $G_{yz}$  \\
$E_y$ & $m2m 1'$ & $Q_y$, $Q^\alpha_y$, $Q^\beta_y$, $G_{zx}$\\
$E_z$ & $mm2 1'$ & $Q_z$, $Q^\alpha_z$, $Q^\beta_z$, $G_{xy}$\\
\hline
$H_x$ & $mm'm'$ & $M_x$, $M^{\alpha}_x$, $M^\beta_{x}$, $T_{yz}$ \\
$H_y$ & $m'mm'$ & $M_y$, $M^{\alpha}_y$, $M^\beta_{y}$, $T_{zx}$ \\
$H_z$ & $m'm'm$ & $M_z$, $M^{\alpha}_z$, $M^\beta_{z}$, $T_{xy}$  \\
\hline
$J_x$ & $m'mm$ & $T_x$, $T^{\alpha}_x$, $T^\beta_{x}$, $M_{yz}$ \\
$J_y$ & $mm'm$ & $T_y$, $T^{\alpha}_y$, $T^\beta_{y}$, $M_{zx}$  \\
$J_z$ & $mmm'$ & $T_z$, $T^{\alpha}_z$, $T^\beta_{z}$, $M_{xy}$   \\
\hline\hline  
\end{tabular}
}
\endgroup
\end{center}
\end{table}

\begin{table}[tb!]
\begin{center}
\caption{
Multipoles induced by $\bm{E}=(E_x, E_y, E_z)$, $\bm{H}=(H_x, H_y, H_z)$, and $\bm{J}=(J_x, J_y, J_z)$ under the monoclinic magnetic point group (MPG) $2/m 1'$. 
We omit the multipoles for the electric monopole, the electric toroidal dipole, and the electric toroidal octupole for simplicity. 
}
\label{tab: C2h}
\begingroup
\renewcommand{\arraystretch}{1.1}
\scalebox{1.0}{
 \begin{tabular}{lcc}
 \hline \hline
$\bm{F}$ & MPG & multipoles \\ \hline
 ---      & $2/m 1'$ & $Q_u$, $Q_v$, $Q_{zx}$ \\
 \hline
$E_x$, $E_z$ & $m 1'$ & $Q_x$, $Q_z$, $Q^\alpha_x$, $Q^\alpha_z$, $Q^\beta_x$, $Q^\beta_z$, $G_{yz}$, $G_{xy}$  \\
$E_y$ & $2 1'$ & $Q_y$, $Q_{xyz}$, $Q^\alpha_y$, $Q^\beta_y$, $G_0$, $G_u$, $G_v$, $G_{zx}$\\
\hline
$H_x$, $H_z$ & $2'/m'$ & $M_{x}$, $M_{z}$, $M^\alpha_x$, $M^\alpha_z$, $M^\beta_x$, $M^\beta_z$, $T_{yz}$, $T_{xy}$ \\
$H_y$ & $2/m$ & $M_y$, $M_{xyz}$, $M^\alpha_y$, $M^\beta_y$, $T_0$, $T_u$, $T_v$, $T_{zx}$ \\
\hline
$J_x$, $J_z$ & $2'/m$ & $T_{x}$, $T_{z}$, $T^\alpha_x$, $T^\alpha_z$, $T^\beta_x$, $T^\beta_z$, $M_{yz}$, $M_{xy}$ \\
$J_y$ & $2/m'$ & $T_y$, $T_{xyz}$, $T^\alpha_y$, $T^\beta_y$, $M_0$, $M_u$, $M_v$, $M_{zx}$  \\
\hline\hline  
\end{tabular}
}
\endgroup
\end{center}
\end{table}

In this section, we classify the induced multipoles by external fields and currents under magnetic point groups. 
We here consider five magnetic point groups, $4/mmm1'$, $6/mmm 1'$, $\bar{3}m1'$, $mmm1'$, and $2/m 1'$, where any of electric quadrupoles $Q_{2m}$ belong to the totally symmetric irreducible representation; the results for other magnetic point groups can be straightforwardly obtained by using the compatible relations between the above groups and subgroups~\cite{Yatsushiro_PhysRevB.104.054412}. 
In each point group, we show the symmetry reduction by the electric field $\bm{E}=(E_x, E_y, E_z)$, the magnetic field $\bm{H}=(H_x, H_y, H_z)$, and the electric current $\bm{J}=(J_x, J_y, J_z)$, and present the induced multipoles. 
The notations of the crystal axes and the multipoles are followed by Ref.~\cite{Yatsushiro_PhysRevB.104.054412}. 
We show the results for the tetragonal magnetic point group $4/mmm 1'$ in Table~\ref{tab: D4h}, the hexagonal magnetic point group $6/mmm 1'$ in Table~\ref{tab: D6h}, the trigonal magnetic point group $\bar{3}m 1'$ in Table~\ref{tab: D3d}, the orthorhombic magnetic point group $mmm 1'$ in Table~\ref{tab: D2h}, and the monoclinic magnetic point group $2/m 1'$ in Table~\ref{tab: C2h}. 

We discuss the result for $4/mmm 1'$ in Table~\ref{tab: D4h}, where the electric quadrupole $Q_u$ belongs to the totally symmetric irreducible representation. 
When the $x$-directional electric field $E_x$ is applied, the electric dipole $Q_x$, the electric quadrupole $Q_v$, the electric octupoles $(Q_x^\alpha, Q_x^\beta)$, and the electric toroidal quadrupole $G_{yz}$ are induced. 
Among them, $(Q_x, G_{yz}, Q_x^\alpha, Q_y^\alpha)$ results from the coupling between $Q_u$ and $E_x$, as shown in Table~\ref{tab: mp}. 
Meanwhile, the remaining $Q_v$ is secondary induced through the symmetry lowering by the breaking of the fourfold rotational symmetry. 
When the $x$-directional magnetic field $H_x$ is applied, the multipoles with opposite $\mathcal{P}$ and $\mathcal{T}$ parities, $(M_x, T_{yz}, M_x^\alpha, M_x^\beta)$, are induced except for $Q_v$. 
Similarly, the $x$-directional electric current $J_x$ induces the multipoles with the opposite $\mathcal{T}$ parity, $(T_x, M_{yz}, T_x^\alpha, T_x^\beta)$, are induced except for $Q_v$. 
These results are consistent with the results in Sec.~\ref{sec: Active multipoles under external fields and currents}. 
In this way, Eqs.~(\ref{eq:Dx})--(\ref{eq:Ozb}) provide useful information to deduce the multipoles by external fields and currents under the electronic nematic orderings with the electric quadrupole.

\section{Summary}
\label{sec: Summary}

We have investigated the nature of the electronic nematic orderings by focusing on the effect of external fields and currents accompanying the breakings of the spatial inversion and/or time-reversal symmetries. 
Based on the multipole representation consisting of four-type multipoles, we showed the conditions to activate rank-1 dipole, rank-2 quadrupole, and rank-3 octupole moments with distinct spatial inversion and time-reversal parities. 
In contrast to the conventional isotropic system, the nematic states characterized by the electric quadrupole exhibit peculiar responses to external stimuli according to the induced unconventional multipoles; the electric toroidal quadrupole, the magnetic toroidal quadrupole, and the magnetic quadrupole are induced by the electric field, magnetic field, and electric current. 
We also summarized the classification of multipoles under several magnetic point groups in a systematic way. 
Our results provide further exploration of characteristic physical phenomena under electronic nematic orderings by external fields and currents.

\begin{acknowledgments}
This research was supported by JSPS KAKENHI Grants Numbers JP21H01037, JP22H04468, JP22H00101, JP22H01183, JP23H04869, JP23K03288, and by JST PRESTO (JPMJPR20L8) and JST CREST (JPMJCR23O4).  
\end{acknowledgments}

\appendix

\bibliographystyle{apsrev}
\bibliography{../ref.bib}
\end{document}